# A Vague Improved Markov Model Approach for Web Page Prediction

Priya Bajaj and Supriya Raheja

Department of Computer Science & Engineering, ITM University
Gurgaon, Haryana 122001, India

## ABSTRACT

*Today most of the information in all areas is available over the web. It increases the web utilization as well as attracts the interest of researchers to improve the effectiveness of web access and web utilization. As the number of web clients gets increased, the bandwidth sharing is performed that decreases the web access efficiency. Web page prefetching improves the effectiveness of web access by availing the next required web page before the user demand. It is an intelligent predictive mining that analyze the user web access history and predict the next page. In this work, vague improved markov model is presented to perform the prediction. In this work, vague rules are suggested to perform the pruning at different levels of markov model. Once the prediction table is generated, the association mining will be implemented to identify the most effective next page. In this paper, an integrated model is suggested to improve the prediction accuracy and effectiveness.*

## 1. INTRODUCTION

Web caching or prefetching is one of the adaptive utility or the approach that analyze the web usage done by the particular user or the users. Based on this analysis, it defines some prediction approach to identify the next expected visiting page before the user demand. When a user is reading his current accessed page, the next predicted page is loaded into the user cache memory. It decreases the loading time for next page access at user end so that the web page retrieval efficiency will be improved.

The concept of web page prediction is the application comes under the web page mining along with data mining. When the page access is performed, it comes under the web content mining to locate and load the predicted page into the cache. When the history of the web server is collected in the form of user web usage history and presented in the form of web pages. The basic attributes of web page history is shown in table 1.

Table 1: Effective Attributes of Web Usage DB

| Attribute | Description |
| --- | --- |
| User IP | Defines the IP address of the web user. |
| Server IP | Defines the IP address of the proxy server the user is associated |
| Access Web Page | Contains the exact file address that user accessed in the form of URL |
| Access Web Domain | Defines the actual web server that represents the domain name |
| Access IP Address | Defines the IP Address of the Web Server |
| Date | Defines the Date, when the page is accessed |
| Time Stamp | Defines the Time, when the page is accessed |

Once the information database gets available, the next work is to perform the data mining operations to prediction. But generally, the size of this kind of datasets is quite large, because of this to reduce the dataset size, some clustering process is required. The clustering can be static session based clustering or an intelligent clustering using some analytical approach. Once the clustering is performed, the identification of the appropriate cluster is performed to that relates the user existence. This identified cluster is selected as the working dataset based on which the prediction is performed.

The prediction process is basically to identify the frequency of next visiting pages in relevancy to the current page. Once the prediction analysis is performed, the association identification is performed to identify most associated next page. This page is then selected as the next predicted web page. The basic structural model of this working process is shown in figure 1.

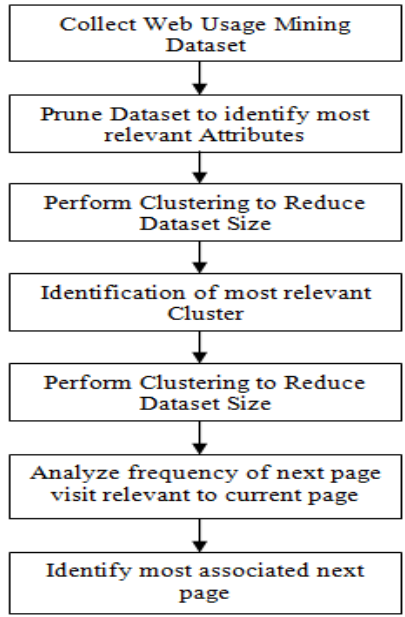

Figure 1 : Basic Structure of Web Page Prediction

In this paper, an improved web page prediction model is presented. The presented work is the improved with the association of three main concepts: markov model, vague rules and the association mining. Markov model will work as the intelligent prediction approach that will be filtered at two different levels using vague rules. Vague will define the intelligent ruleset by performing the dataset analysis. At the later stage, the association mining will be implemented to perform the web page prediction for the caching.

In this section, an introduction to the web page prediction is defined with the specification of the mining dataset and the structure. The structure is the most effective process steps followed by most of the researchers. In section II, the work done by the earlier researchers for web page prediction is discussed. In section III, the exploration of proposed vague improved markov model is defined. In section IV, the conclusion obtained from the work is presented.

## 2. LITERATURE REVIEW

Lot of work is already done in the area of web page prediction and web caching. In this section, the work done by the earlier researchers in this area is presented and discussed.

In this paper, author has defined the optimization process to reduce the web information access and to reduce the error. Author optimized the search mechanism along with encoded search. Author improved the quality of the search algorithm with the reduction of integration error[1]. Another work on the improvement of web page access was defined by the author. Author presented the prediction analysis approach to improve the web page caching. The work proposed by the author considered a realistic prefetching architecture using real and representative traces. Author implemented the work in real web environment and the obtained results shows the significant improvement over the existing approaches [2]. In this paper Author present an online prediction model that does not have an offline component and fit in the memory with good prediction accuracy. Presented algorithm is based on LZ78 and LZW algorithms that are adapted for modeling the user navigation in Web. Presented model decreases computational complexities which are a serious problem in developing online prediction systems. A performance evaluation is presented using real Web logs. This evaluation shows that Presented model needs much less memory than PPM family of algorithms with good prediction accuracy [3].

In this paper preliminary work in the area of Web page prediction is presented. The designed and implemented prototype offers personalized interaction by predicting the user's behavior from previous Web browsing history. Those predictions are afterwards used to simplify the user's future interactions. Rather simple and feasible prototype enhancements are offered and discussed. Its simplicity and effectiveness makes it potentially useful for widespread application [4]. In this paper, Author presented an improvement over the caching scheme so that the page access consistency will be improved. Author performed the analysis over the cache parameters in terms of size, frequency analysis etc so that effective web page modeling under prefetching will be done [5]. In this paper three different schemes for Web Prefetching and caching are proposed i.e. Prefetching only, Prefetching with Caching and Prefetching from Caching. Prediction of the next accessed Web page for prefetching and caching is achieved by modeling the Web log using Dynamic Nested Markov model. Dynamic Nested Markov model is analyzed on these three Prefetching and Caching schemes. Experiments have been conducted on real world data sets [6].

**V.V.R.Maheswara Rao** defined a Markov Prediction Model called HSMP. The HSMP model is initially predicts the possible wanted categories using Relevance factor, which can be used to infer the users' browsing behavior between Web categories. Then predict the pages in predicted categories using techniques for intelligently combining different order Markov models so that the resulting model has low state complexity, improved prediction accuracy and retains the coverage

International Journal of Computer Science & Engineering Survey (IJCSES) Vol.5, No.2, April 2014

of the all higher order Markov model [7]. R.Khanchana defined an approach that uses a HMM model to perform the single page analysis. Author uses a directed graph based weightage analysis approach to identify the integration links between the web pages. Author generated the navigation path to reduce the size of processing dataset and to perform accurate prediction of the web page. The obtained results from the system gives the effective ranking the web page in terms of rank assignment and the page prediction [8]. Naveed Ahmad focused on when a user requests for a Web page, how to improve the overall performance of Web prefetching mechanism? The proposed mechanism provides the pages locally available to a user or group of users by utilizing bandwidth of the network. The server contains an algorithm for the prediction of Web pages and the prediction of a Web page is based on counting the number of times a page is accessed by a user from each cluster [9].

Yaser Alosefer has presented an algorithm that is able to detect the potential malicious behavior of a Web server based on current and past interactions between the Web client and the server and can also predict possible future behaviors. The prediction algorithm learns from previously scanned behaviors recorded by a client honeypot system. Author group such behaviors in order to enable common characteristics to be investigated across these groups [10]. Shreya Dubey has defined a hybrid model for the Web page prediction. This model includes some intelligent approaches called SBM, Association rule mining and the Markov model to perform the page prediction. The author combined two main concepts called the prediction mining and the Web personalization [11]. Sina Bahram has defined a work on the prediction of Web pages under the machine learning approaches. Author defined the structural and featured analysis on the Web pages to identify the individual and the relation features over the Web access. Author has defined three main datasets to perform the classification process. Author implemented the work in real environment and obtained results shows the effectiveness of the work [12].

## 3.PROPOSED WORK

The presented work is the improvement of the existing prediction model with an effective prediction model with rule based pruning process. The presented process is effectual for large dataset as the work includes four levels of filtration process. This filtration process includes the static session based filtration, dynamic clustering process and two layers of pruning process. The complete work is divided in three main stages called filtration stage, Analysis stage and Prediction Stage. The basic structure of proposed architecture is shown in figure 2.

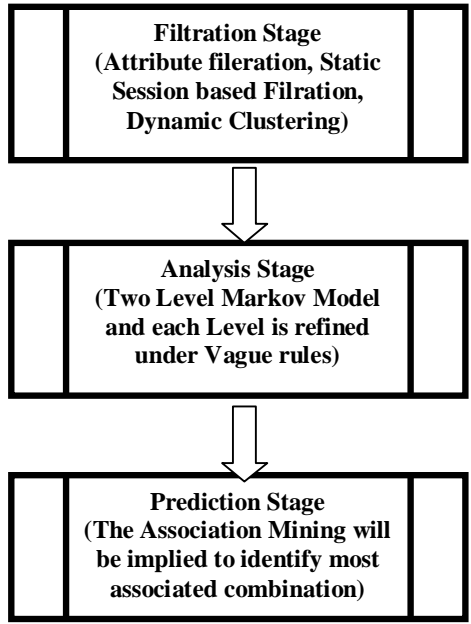

Figure 2 : Structure of Proposed Work

## 3.1 Filtration Stage

The filtration is actually the identification stage of most relevant dataset on which the actually prediction and analysis process will be performed. When the raw web data is collected, it contains number of attributes and having a large tupleset. It is not feasible to process on complete dataset at one time. Because of this, the filtration stage is implied over it perform the dataset reduction. This dataset reduction process includes the horizontal and vertical filtration. The horizontal dataset reduction includes the elimination of non required attributes from the dataset and identifies the most relevant attributes. The vertical dataset reduction process includes the static and dynamic dataset reduction process. The static dataset reduction is defined in terms of session based reduction. The session can be defined in terms of time line or the proxy server. In dynamic dataset reduction is actually the clustering process that will collect the most relevant pages. The filtration stage process is here shown in figure 3.

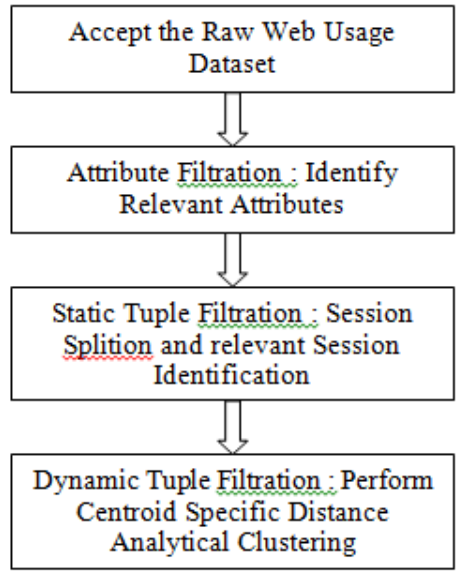

Figure 3 : Filtration Process

This filtration stage actually identifies the most relevant sub-dataset on which the actual prediction process will be performed. This stage will improve the accuracy and efficiency in the prediction of end result.

### 3.2 Analysis Stage

This is actually the process stage defined in the paper. The formation of this stage includes an integrated combination of markov model with vague rule set. In this stage, most relevant cluster will be accepted as the input dataset on which the markov model will be implemented at two level . In level 1, the single page analysis will be performed under the frequency analysis.

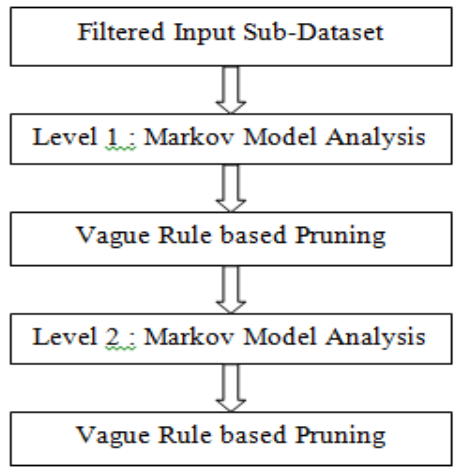

Figure 4 : Analysis Process

This stage will be followed by the vague ruleset to perform the dataset pruning so that the irrelevant and less frequency pages will be eliminated from the list. After the pruning process, the

54

level two analysis will be performed under markov model to perform the associated web page analysis in combination of two pages. This analysis again include the identification of the associatively and the frequency of associated page combination. This level 2 analysis will be followed by the pruning process defined using vague ruleset. The model of this analysis stage is shown in figure 4.

### 3.3 Prediction Stage

This stage is actually the conclusion stage where the identification of the most associated page will be performed. In this stage, the input will be taken as the most effective page pair combination driven from the anlaysis stage. Now, the association mining will be implemented on this anlytical dataset to obtain the effective result. The most associated page combination will be elected as the final cached page.

### 4. CONCLUSION

In this present work, an effective page prediction model is presented using vague improved markov model. This paper has presented the conceptual model of the presented work with detailed exploration of each stage.

**Authors**

**First Author Priya Bajaj**, Have completed engineering in Computer Science from Maharishi Dayanand University, Rohtak in 2012 and pursuing M-tech in Computer Science from ITM University (2012-2014).

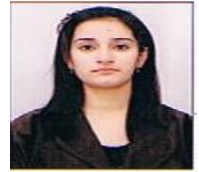

**Second Author Supriya Raheja, Assistant Professor**, ITM University, is pursuing her PhD in Computer Science from Banasthali University. She had done her engineering from Hindu college of Engineering, Sonepat and masters from Guru Jambeshwar University of Science and Technology, Hisar. Her total Research publications are thirteen in International Conferences and Journals. She is working as a Reviewer/Committee member of various International Journals and Conferences.

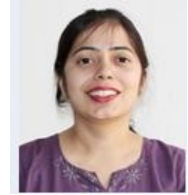